\documentstyle[epsfig,aps,floats,eqsecnum,preprint]{revtex}

\def \tablerule{\noalign {\vskip3truept\hrule\vskip3truept}}
\def\laq{\raise 0.4ex\hbox{$<$}\kern -0.8em\lower 0.62
ex\hbox{$\sim$}}
\def\gaq{\raise 0.4ex\hbox{$>$}\kern -0.7em\lower 0.62
ex\hbox{$\sim$}}

\newcommand{\beq}{\begin{equation}}
\newcommand{\eeq}{\end{equation}}
\newcommand{\be}{\begin{equation}}
\newcommand{\ee}{\end{equation}}
\newcommand{\bea}{\begin{eqnarray}}
\newcommand{\eea}{\end{eqnarray}}
\newcommand{\pa}{\partial}

\newcommand{\vrsmall}{\vrule width 0pt height 18pt depth 15pt}

\begin{document}
\draft
\preprint{\vbox{\baselineskip=12pt
\rightline{CERN-TH/97-213}
\vskip0.2truecm
\rightline{hep-th/llmmnnn}}}

\title{\Large\bf Quantum Inhomogeneities in String Cosmology}
\author{A. Buonanno\footnote{Present address:
Institut des Hautes Etudes Scientifiques, 91440 Bures-sur-Yvette, France.},
K.A. Meissner\footnote{Permanent address: Institute for Theoretical Physics,
Ho{\.z}a 69, 00-689 Warsaw, Poland.},
C. Ungarelli\footnote{Present address: Dipartimento di Fisica,
Universit\`a di Pisa, Piazza Torricelli 2, 56100 Pisa, Italy.}
and G. Veneziano}
\address{Theory Division, CERN, CH-1211 Geneva 23, Switzerland}
\maketitle
\begin{abstract}
Within two specific string cosmology scenarios --differing in the way the
pre- and post-big bang phases are joined-- we compute
the size and spectral slope
of various types of cosmologically amplified quantum fluctuations
that arise in generic compactifications of  heterotic string theory.
By further imposing  that these perturbations become the dominant source
of energy at the onset of the radiation era, we obtain physical bounds on
the background's moduli, and discuss the conditions under which both  a
(quasi-) scale-invariant spectrum of axionic perturbations  and
sufficiently large  seeds for the galactic magnetic fields are generated.
We also point out a potential problem with achieving the exit to the
radiation
era when the string coupling is near its present value.
\end{abstract}
\vskip 0.3truecm
\pacs{\tt PACS number(s): 04.50.+h, 98.80.Cq}

\section {Introduction}
\label{sec1}
Perhaps the most appealing feature of standard
inflationary cosmology~\cite{infl} is its ability
to stretch out generic/arbitrary initial classical inhomogeneities and
to replace them by a calculable spectrum of cosmologically
amplified quantum fluctuations. The latter behave, for all physical
purposes, as
a set of properly normalized stochastic classical perturbations.
A much advertised outcome of slow-roll inflation
is a (quasi-) scale-invariant (Harrison-Zeldovich (HZ))
spectrum of density fluctuations, a highly desirable feature
for explaining both the CMB temperature fluctuations on large angular scales
and the large-scale structure of the visible part of our Universe.

The so-called pre-big bang (PBB) scenario~\cite{MG1,MG2} offers,
within the context of string theory, an
alternative to the usual inflationary paradigm.
Provided a graceful exit can be achieved~ (see \cite{GMV,BM} for recent
progress on this issue),
the PBB scenario exhibits  several appealing advantages, e.g.
\begin{list}{$\bullet$}{\setlength{\leftmargin}{.5in} 
\setlength{\rightmargin}{.5in}}
\item it naturally provides  inflationary solutions
through the duality symmetries~\cite{duality} of string theory;
\item
it assumes a natural, simple, initial state for the Universe, which is fully
under control: the perturbative vacuum of superstring theory;
\item
it needs no fine-tuning of couplings and/or potentials: the
inflaton is identified with the
dilaton, which is ubiquitous in string theory, is effectively massless
at weak coupling, and provides inflation through its kinetic  energy;
\item
it can provide a hot big bang initial state  as a late-time
{\it outcome} of the pre-big bang phase, through the
amplification of vacuum quantum fluctuations generated in this latter phase.
\end{list}

In recent work~\cite{inh,BMUV} we have discussed the conditions under which
classical inhomogeneities get efficiently erased
in string cosmology. In general, this does occur provided two moduli of the
classical solutions at weak coupling and curvature (basically an initial
coupling and an initial curvature scale) are bounded from above.
Whether such conditions correspond to an acceptable degree of
fine-tuning of
the initial conditions or not is still the matter of some
controversy~\cite{TW,MaS,BMUV}.

An interesting outcome of these investigations has been a
motivated conjecture \cite{BMUV} that, for negative spatial
curvature, the pre-big bang
phase  itself
is generically  preceded by a contracting ``Milne" phase,
corresponding to a particular parametrization of the past light cone
of trivial Minkowski space-time with a constant dilaton. Such a
background, the trivial all-order classical vacuum of superstring
theory, turns out to be an unstable early-time
fixed point of the evolution. Thanks to dilaton/metric fluctuations, it
appears to lead, inevitably, to pre-big bang-type
inflation at later times.

In this paper we shall assume that the above classical picture
effectively wipes out, during its long pre-big bang phase,
spatial curvature and classical
inhomogeneities, and we move on to analyse the second alleged virtue of
inflationary
cosmology, the generation of an interesting spectrum of amplified quantum
fluctuations. As several previous investigations have shown
\cite{tensor,scalar,em},
achieving this
is not at all automatic in string cosmology.
It was soon realized that, in the simplest PBB scenario, tensor
~\cite{tensor} and scalar-dilaton~\cite{scalar} perturbations tend to
have steep spectra (typically a spectral index $n=4$, as compared to
HZ's $n=1$). Perturbations of gauge fields coming from compactification of
the extra 16 bosonic dimensions of heterotic string theory can
have somewhat smaller spectral indices~\cite{em}, but still in the range
$3<n<4$.

The situation can be improved by assuming~\cite{tensor} that a
long string phase (during which the dilaton grew linearly in cosmic time
while the Universe expanded exponentially) took place between the
dilaton and the usual FRW phase. In such a case, it is possible to get
{\it either}
an interesting spectrum of gravitational waves~\cite{tensor} in the
range of interest for
detection, {\it or} enough EM perturbations to explain the magnetic
fields~\cite{em},
but  not both, apparently. A flat spectrum of EM perturbations, which
can possibly provide a new mechanism for generating large scale
structure~\cite{em1}
is not excluded either.

Recently, however, Copeland et al.~\cite{Copeland} made the
interesting observation that
axionic perturbations, even in the absence of a string phase, can have a
flat spectrum, depending on how the internal dimensions evolve during the
dilatonic phase.
Unfortunately, Copeland et al. stopped short of computing the axionic
spectrum after re-entry. Nonetheless, their result hints at a possible
dominance of axionic perturbations over all others and calls for a revision
of the whole scenario and of the phenomenological constraints that
must be imposed on it.

In this paper we analyse quantum fluctuations
of various kinds in two distinct  scenarios for the background, with or
without an intermediate string phase. We may expect either
possibility to occur, depending on the precise mechanism
 providing
the transition (exit) from the PBB phase into the FRW phase.

An intermediate string phase is  natural
if we assume \cite{GMV} that $\alpha'$ corrections provide a
non-perturbative fixed point with a high constant curvature and a linearly
growing dilaton. In this case
we expect the transition to the FRW phase to occur during the string
phase as soon as the energy stored in
the quantum fluctuations reaches criticality (recall that the
condition of criticality
depends on the coupling). This is like saying that the final
transition to the radiation-dominated era will be induced by
string-loop,~back-reaction effects (see, e.g. \cite{BM}).

We can imagine, instead, that $\alpha'$ corrections are sufficient
 to provide by themselves
a sudden branch-change from the perturbative PBB phase to another
duality-related vacuum phase, with the Hubble parameter
making a bounce around its maximal value. In the language of~\cite{GMV}
this would
correspond to a square-root-type vanishing of a $\beta$-function.
Again, the dual ($-$ branch)
phase will gently yield  to a FRW Universe as soon as the energy
stored in the quantum fluctuations becomes critical.

As already mentioned, an important ingredient of our approach
 is the (self-consistency) requirement of criticality
at the beginning of the radiation era. This provides a new relation between
the moduli of the PBB background and the coupling and energy density
(or temperature) at the beginning of the radiation era.
 As we will see,
 the dilaton at the beginning of this era is generically displaced  from
its eventual/present value; hence  this
primordial radiation era is not yet quite the one of standard cosmology.
 It may take a while before the non-perturbative
dilaton potential makes its presence felt and forces the dilaton to
its minimum.
The detailed study of such post-big bang phase is left to further work.

One of the main conclusions of this paper is  that, provided
$U(1)_{em}$  has a component in the Kaluza-Klein gauge group
produced in the compactification from $D=10$ to $D=4$, sufficiently large
seeds for galactic magnetic fields can be generated, even in the absence of
a string
phase. Furthermore, this happens in the same range of moduli
for which a nearly scale-invariant spectrum of axionic perturbations is
generated. Such a range includes a particularly symmetric point in moduli
space, the one corresponding to isotropy (up to T-duality
transformations) in
all nine spatial dimensions.

The outline of the paper is as follows: In Sec.~\ref{sec2} we
give, for the sake of completeness, the four-dimensional low-energy
string-level heterotic effective action that we will work
with. In Sec.~\ref{sec3} we fix our parametrization of the backgrounds
for the two previously discussed scenarios. In Sec.~\ref{sec4} we
 derive general formulae
for the spectra of various perturbations,
which get amplified by
a generic background of the kind discussed in Sec.~\ref{sec3}.
We will verify that
our spectra satisfy a  ``duality" symmetry that
can be shown to follow from  general arguments~\cite{dualitypert}.
In Sec.~\ref{sec5} we give the explicit form of the spectra for the
two backgrounds discussed in Sec.~\ref{sec3},  and present them in
various tables and plots.
We will also impose the criticality condition and
discuss its immediate consequences. Finally, Sec.~VI  contains a discussion
of the results and some  conclusions.

This paper is somewhat technical in nature and contains
explicit general formulae that can be useful to the practitioner but
do not carry  easy messages: these
 can be better found in the tables and  figures.
At any rate, in order to help the reader, we have  relegated
the most complicated formulae to an appendix.

\section{ String effective action from dimensional reduction}
\label{sec2}
Following the notations of~\cite{MS} we consider superstring theory
in a space-time
${\cal M} \times {\cal K}$, where ${\cal M}$, with Minkowskian signature,
 has four non-compact dimensions, and
${\cal K}$ consists of six compact dimensions upon which
 all fields are assumed to be independent.
Local coordinates of ${\cal M}$ are labelled by $\mu, \nu, \rho = 0,
\dots ,3$, those
of ${\cal K}$ by $a, b, c = 4, \dots ,9$. Moreover, all ten-dimensional
fields  and indices are distinguished by a hat.

We will limit ourselves to the case of a diagonal  metric for the
internal six-dimensional compact space,
of a non-vanishing internal antisymmetric-tensor
and of one Abelian heterotic $U(1)$ gauge field $A_\mu$:
\begin{equation}
\hat{g}_{\hat{\mu} \hat{\nu}}= \left ( \begin{array}{cc}
g_{\mu \nu} + g_{a b}\,V_{\mu}^{\,\,a}\,V_{\nu}^{\,\,b} &
g_{a b}\,V_{\mu}^{\,\,b} \\
g_{a b}\,V_{\nu}^{\,\,b} & g_{a b}\\
\end{array}
\right )\,,
\end{equation}
\begin{equation}
\hat{B}_{\hat{\mu} \hat{\nu}}= \left ( \begin{array}{cc}
B_{\mu \nu} & W_{\mu a} - B_{a b}\,V_{\mu}^{\,\,b} \\
-W_{\nu a} + B_{a b}\,V_{\nu}^{\,\,b} & B_{a b}\\
\end{array}
\right )\,.
\end{equation}
In the following we take $g_{a b} = e^{2\sigma_a}\,\delta_{a b}$.

The low-energy four-dimensional effective string action is
\bea
&& {\cal S}^B_{\rm eff} = \frac{1}{2\lambda_s^{2}}\, \int
d^4x\,\sqrt{-g}\,e^{-\varphi}\,\left [
{\cal R} + g^{\mu \nu}\,\pa_\mu\varphi\,\pa_\nu\varphi- \,g^{\mu \nu}\,
\pa_\mu \sigma_a\,\pa_\nu\sigma_a -
\frac{1}{4}\,e^{2\sigma_a}\,V_{\mu \nu}^{\,\,\,\,a}\,V^{\mu \nu a}
\right . \nonumber \\
&& \left . - \frac{1}{4}\, e^{-2 \sigma_a}\,H_{\mu \nu a}\,H^{\mu \nu
}_{\,\,\,\,\,a}
-\frac{1}{12}\,H_{\mu \nu \rho}\,H^{\mu \nu \rho} +
\frac{1}{4} F_{\mu \nu}\,F^{\mu \nu} -\frac{1}{4}\,g^{\mu \nu}
e^{-2\sigma_b}\,e^{-2\sigma_c}\,\partial_\mu B_{b c}\, \partial_\nu B_{b c}
\right ]\,,
\eea
where $\lambda_s = \sqrt{8\pi}/M_s$ is the string-length parameter,
\beq
\label{B}
H_{\mu \nu \rho} = \pa_\mu B_{\nu \rho}
- \frac{1}{2}\, \left [ V_\mu^{\,\,a} \,W_{\nu \rho a} +
W_{\mu a} \,V_{\nu \rho}^{\,\,\,\,a} \right ]
- \frac{1}{2}A_\mu\,F_{\nu \rho} + \quad \mbox{cyclic perm.}\,,
\eeq
\bea
&& H_{\mu \nu a} = W_{\mu \nu a} - B_{a b}\,V_{\mu \nu}^{\,\,\,\,b}\,,
\quad \quad F_{\mu \nu } = \partial_\mu A_{\nu} -\partial_\nu A_{\mu} \,, \\
&& V_{\mu \nu}^{\,\,\,\,a} = \partial_\mu V_\nu^{\,\,a} -\partial_\nu
V_\mu^{\,\,a} \,,
\quad \quad W_{\mu \nu a} = \partial_\mu W_{\nu a} -\partial_\nu
W_{\mu a} \,,
\eea
and $\varphi$  stands for the effective four-dimensional dilaton field:$$
\varphi = \phi  -\sum_a \sigma_a\,.
$$
The components of the antisymmetric tensor $H^{\mu \nu \rho}$ with
$\mu, \nu, \rho = 0,1,2,3$ can  be rewritten in
terms of the pseudoscalar axion $A$ as
\beq
H^{\mu \nu \rho} \equiv E^{\mu \nu \rho \sigma} \,e^\varphi\,\pa_\sigma A\,,
\eeq
where $E^{\mu \nu \rho \sigma}$ is the covariant full antisymmetric
Levi-Civita tensor, which satisfies ${\cal D}_\alpha  E^{\mu \nu \rho
\sigma} = 0$.
Using Eq.~(\ref{B}), and imposing the Bianchi identity ($d^2 B = 0$),
we get the equation of motion for the axion field
\beq
\pa_\mu(e^\varphi\,\sqrt{-g}\,g^{\mu \nu}\,\pa_\nu A) -
\frac{1}{8}\,\frac{\epsilon^{\mu \nu \rho \sigma}}{\sqrt{-g}}\,
\left [ 2 W_{\mu \nu a}\,V_{\rho \lambda}^{\,\,\,\,a} +
F_{\mu\nu}\,F_{\rho \sigma}\right ]  =0 \,.
\eeq
The reduced action then becomes
\bea
\label{redaction}
&& {\cal S}^A_{\rm eff} = \frac{1}{2\lambda_s^{2}}\, \int
d^4x\,\sqrt{-g}\,e^{-\varphi}\,\left [
{\cal R} + g^{\mu \nu}\,\pa_\mu\varphi\,\pa_\nu\varphi- \,g^{\mu \nu}\,
\pa_\mu \sigma_a\,\pa_\nu\sigma_a -
\frac{1}{4}\,e^{2\sigma_a}\,V_{\mu \nu}^{\,\,\,\,a}\,V^{\mu \nu a}
\right . \nonumber \\
&& \left . - \frac{1}{4}\, e^{-2 \sigma_a}\,H_{\mu \nu a}\,
H^{\mu \nu }_{\,\,\,\,\,a}
-\frac{1}{2}\,e^{2 \varphi}\,g^{\mu \nu}\,\pa_{\mu}A\,\pa_\nu A
- \frac{1}{8}\,e^{\varphi}\,\frac{A\,\epsilon^{\mu \nu \rho \sigma}}
{\sqrt{-g}}\,
\left [ 2 W_{\mu \nu a}\,V_{\rho \lambda}^{\,\,\,\,a} + F_{\mu \nu}\,
F_{\rho \sigma} \right ]\right . \nonumber \\
&& \left. + \frac{1}{4} F_{\mu \nu}\,F^{\mu \nu} -\frac{1}{4}\,g^{\mu \nu}
e^{-2\sigma_b}\,e^{-2\sigma_c}\,\partial_\mu B_{b c}\,\partial_\nu
B_{b c} \right ]\,.
\eea
We are interested in fluctuations around a homogeneous background
with $\varphi = \varphi(t)$,
$$
g_{\mu \nu} = (-1,a^2(t)\,\delta_{i j},
b_c^2(t)\,\delta_{c d}) \,,\quad \quad i,j = 1,2,3 \,, \quad \quad
c,d=4,\dots ,9 \,,
\quad \quad \mbox{\rm all other fields} =0\,.
$$
In the following we will also use the metric
$g_{\mu \nu} = (-a^2(\eta),a^2(\eta) \delta_{i j},
b_c^2(\eta)\,\delta_{d c})$, where
we have introduced the conformal time $\eta$ by
 $d \eta = d t/a$.

\section{Two models for the background}
\label{sec3}
If the initial value of the string coupling
is sufficiently small, it is possible for the Universe to reach the
high curvature regime, where  higher-derivative corrections are
important,
while the string coupling is still small enough to neglect loop
corrections ($g=\exp \varphi/2 \ll 1$).
As discussed in the introduction, we will consider two extreme
alternatives. In the first,
 $\alpha^\prime$ corrections ``lock" the Universe  in a {\it string phase}
with a constant $H$ and a linearly growing
 dilaton (with respect to cosmic time) \cite{GMV}; in the second
scenario, $\alpha^\prime$ corrections
induce a sudden transition from a perturbative $(+)$ branch solution to a
perturbative $(-)$ branch phase. We will refer to the latter
as the {\it dual-dilaton phase}.

We will thus consider a PBB cosmological background in which
the Universe starts in the perturbative string vacuum, reaches the string
curvature scale while in a dilaton-vacuum solution,
goes either to the dual-dilaton phase or to the string phase, and
finally  enters the radiation
era as a result of the back-reaction from the amplified quantum
fluctuations.
We now parametrize these two scenarios for the backgrounds,
imposing the continuity of $a,a^\prime$, $b_a,b_a^\prime$,
$\varphi$.

\subsection{\bf Intermediate dual dilaton phase }
\label{sec3.1}
\newcounter{spacerule}
\begin{list}{\textbf{\arabic{spacerule}}}{\setlength{\leftmargin}{0in} 
\setlength{\rightmargin}{0in}\usecounter{spacerule}} 
\item \underline{\sl Dilaton phase}

For $-\infty <\eta <\eta_s$, with $\eta_s <0$ we have
\bea
\label{dil}
a(\eta )&=&-\frac{1}{H_s\,\eta_s}\,
\left|\frac{\eta\,(1-\delta) -\eta_s}
{\delta\,\eta_s}\right |^{\delta/(1-\delta)}\,,\\
b_a(\eta) &=& -{H_s\,\eta_s}\,
\left|\frac{(\eta-\eta_s)\,(1-\delta)-\beta_a\,\eta_s}
{\beta_a\,\eta_s}\right |^{\beta_a/(1-\delta)} = e^{\sigma_a} \,,\\
\varphi (\eta ) &=& \varphi_s + \frac{3 \delta -1}{1-\delta}
\log \left |\frac{\eta\,(1-\delta)-\eta_s}
{\delta\,\eta_s} \right | \,, \\
  1 &=& 3\delta^2 + \sum_a \beta_a^2 \,,
\eea
where $H_s = a^\prime(\eta_s)/a^2(\eta_s)$  is of  order  $M_s$. We will
consider the case $\delta <0$ and $\beta_a>0$, i.e. a
superinflationary solution with
 contracting internal dimensions.
Because of the constraint between $\delta$ and  $\beta_a$,
 if $|\delta|<1/\sqrt{3}$, some of the  $\beta_a$ must be non-vanishing.
In what follows we will pick two extreme cases: i) the most isotropic
case, with
 $\beta_a = \sqrt{(1-3\delta^2)/6}$, or ii) the most anisotropic
one, with
$\beta_a = \delta_{a1} \sqrt{(1-3\delta^2)}$. In figures
we shall denote these two cases by a subscript $6$ and $1$, respectively.

\item \underline{\sl Dual-dilaton phase}

For $\eta_s <\eta <\eta_1$, with $\eta_1 >0$ we take
\bea
\label{dual}
a(\eta )&=&-\frac{1}{H_s\,\eta_s}\,
\left|\frac{\eta\,(1-\theta)-\eta_s}
{\theta\,\eta_s}\right |^{\theta/(1-\theta)}\,,\\
b_a(\eta) &=& -{H_s\,\eta_s}\,
\left|\frac{(\eta-\eta_s)\,(1-\theta)-\xi_a\,\eta_s}
{\xi_a\,\eta_s}\right |^{\xi_a/(1-\theta)}\,,\\
\varphi (\eta ) &=& \varphi_s + \frac{3 \theta -1}{1-\theta}
\log \left |\frac{\eta\,(1-\theta)-\eta_s}
{\theta\,\eta_s} \right | \,,
\eea
where $3\theta^2 + \sum_a \xi_a^2 =1$ and we will
fix $\theta >0$ and $\xi_a<0$, i.e. a decelerated expansion for the
external scale factor
and a decelerated contraction for the internal ones. Again, we
distinguish
two cases, $\xi_a = - \sqrt{(1-3\theta^2)/6}$ or $\xi_a = -
\delta_{a1} \sqrt{(1-3\theta^2)}$.
\item \underline{\sl Radiation phase}

In the region $\eta_1 <\eta <\eta_{\rm eq}$, with
$\eta_{\rm eq}$ the time of equivalence between radiation and matter
density,
we write
\bea
\label{rad}
a(\eta )&=&\frac{(\eta-\eta_1-\eta_s)}{H_s\,\eta^2_s}\,
\left|\frac{\eta_1\,(1-\theta)-\eta_s}
{\theta\,\eta_s}\right |^{\theta/(1-\theta)}\,,\\
\varphi (\eta ) &=& \mbox{\rm const.}\,,\quad \quad b_a(\eta) =
\mbox{\rm const.}\,
\eea
\end{list}

\subsection{\bf Intermediate string phase }
\label{sec3.2}
\begin{list}{\textbf{\arabic{spacerule}}}{\setlength{\leftmargin}{0in} 
\setlength{\rightmargin}{0in}\usecounter{spacerule}} 
\item \underline{\sl Dilaton phase}

We parametrize this phase  exactly  as before.
Thus, for $-\infty <\eta <\eta_s <0$, Eqs. (\ref{dil}) to (3.4) hold.

\item \underline{\sl String phase}

For $\eta_s <\eta <\eta_1$, with $\eta_1 <0$
\beq
\label{duals}
a(\eta )=-\frac{1}{H_s\,\eta}\,,\quad \quad b_a(\eta) = \mbox{\rm const.}\,,
\quad \quad
\varphi (\eta )=\varphi_s -2\zeta\,\log
\left(\frac{\eta}{\eta_s}\right ) \,,
\eeq
hence a constant Hubble parameter for the external scale factor.

\item \underline{\sl Radiation phase}

In the range $\eta_1 <\eta <\eta_{\rm eq}$ we have
\beq
\label{rads}
a(\eta )=\frac{1}{H_s\,\eta^2_1}\,(\eta-2\eta_1)\,, \quad \quad b_a(\eta) =
\mbox{\rm const.}\,, \quad \quad  \varphi (\eta ) = \mbox{\rm const.}\,
\eeq
\end{list}

An important property of these backgrounds
is that the derivative of the field $\varphi$ is {\it not} continuous across
the two transitions.
This reflects the fact that we do not have as yet a satisfactory
model for the
transitions from one epoch to another. As discussed below,
this discontinuity creates a  technical
problem, which has to be  judiciously solved in order to correctly compute
the spectrum of perturbations around this kind of backgrounds.

\section{Amplification of vacuum fluctuations}
\label{sec4}
Let us consider a generic massless field, whose quadratic
fluctuations are described by the action
\beq
\label{lag}
\delta {\cal S} = \int d \eta\,\tilde{a}^2\,\left [ (\Psi^\prime)^2 -
(\nabla \Psi)^2 \right ]\,,
\eeq
where a prime stands for derivative with respect to conformal time
$\eta$ and
$\tilde{a}$,  the so-called ``pump" field, is a
homogeneous background field that depends on the particular perturbation
under study.

The safest way to analyse the amplification of the
vacuum fluctuations of  $\Psi$ makes use of a
canonical Hamiltonian approach  and leads
to the derivation
\cite{dualitypert} of certain duality symmetries of the spectra.
We will use instead the simpler Lagrangian method and fix some
ambiguity encountered
in that approach by demanding agreement with the  Hamiltonian treatment.
We believe, of course, that our prescription can also be fully justified
within the Lagrangian framework.

The equation of motion for the Fourier components of $\Psi$ is
\beq
\label{eq}
\Psi_k^{\prime \prime} + 2\,\frac{{\tilde{a}}^\prime}{\tilde{a}}\,
\Psi_k^\prime + k^2\,\Psi_k = 0 \,.
\eeq
Introducing the canonical variable
\beq
v_k = \tilde{a}\, \Psi_k \, ,
\eeq
Eq.~(\ref{eq}) can be rewritten in the form
\be
\label{sch}
\frac{d^2 v_k }{d\eta^2} + \left [k^2 - U(\eta)\right ]\,v_k = 0\,,\ \
\ \ \ U(\eta) = \frac{1}{\tilde{a}}\,\frac{d^2 \tilde{a}}{d \eta^2}\,.
\ee
In order to get general formulae for the spectrum we parametrize the
pump field in the three epochs as follows\footnote{
In order to simplify the final expression of the Bogoliubov
coefficients,
we have slightly changed the constant parameters appearing in the
pump field in the three phases.
With the original parameters of Sect. 3, the Bogoliubov coefficients
would just change by
numerical factors  $O(1)$, but the spectral slopes
and  the ``duality'' symmetry (see sect.~\ref{sec4.2}) 
would still be  the same.}
\bea
\label{apar1}
&& \tilde{a} = \left (\frac{\eta}{\eta_s} \right )^{\gamma}\quad \quad
\quad\quad\quad\,\,\,\,\,\,\,\,\,\,\,\, -\infty < \eta < \eta_s < 0 \,,\\
\label{apar2}
&& \tilde{a} = \left |\frac{\eta-2\eta_s}{\eta_s} \right
|^{\kappa}\quad \quad\quad\quad\quad\,\,\,\,
\eta_s< \eta < \eta_1 \,, \,\,\eta_1 > 0 \,,\\
\label{apar3}
&& \tilde{a} = \left (\frac{\eta}{\eta_1} \right )^{\alpha}\,
\left |\frac{\eta_1-2\eta_s}{\eta_s} \right |^{\kappa}
\quad \quad \eta_1 < \eta < \eta_{\rm eq}\,,
\eea

\subsection{Analytic form for the Bogoliubov coefficients}
\label{sec4.1}
The solutions of the equation of motion (\ref{sch}) for the pump
fields (\ref{apar1}), (\ref{apar2}) and (\ref{apar3}) are respectively
\bea
\label{nor}
&& v_k(\eta) = \sqrt{|\eta|}\,C\,H_\nu^{(1)}(k|\eta|)\,,\\
&& v_k(\eta) = \sqrt{|(\eta-2\eta_s)|}\,\left [
A_+\,H_\mu^{(1)}(k|(\eta-2\eta_s)|) +
A_-\,H_\mu^{(2)}(k|(\eta-2\eta_s)|)\right ]\,,\\
&& v_k(\eta) = \sqrt{|\eta|}\,\left [ B_+\,H_\rho^{(1)}(k|\eta|) +
B_-\,H_\rho^{(2)}(k|\eta|)\right ]\,,
\eea
where
\be
\label{defmu}
\nu= |\gamma-1/2|\,, \ \ \ \mu= |\kappa-1/2|\,,\ \ \ \ \rho=|\alpha-1/2|
\ee
and we have normalized (\ref{nor}) allowing only positive frequencies
in the flat vacuum state at $\eta \rightarrow -\infty$, so that
\beq
v_k(\eta)\to \frac{C}{\sqrt{k}}e^{-ik\eta}
\eeq
and $|C|= 1$. For reasons explained below we impose the continuity of
$\Psi_k$ and of
its first derivative at $\eta_s$, $\eta_1$, {\it not} the continuity
of the canonical field $v_k$. Using the relation
$$H^{(2)\prime}_\mu(z)\,H_\mu^{(1)}(z) -
H^{(1)\prime}_\mu(z)\,H_\mu^{(2)}(z)= - 4i/(\pi z)\,,
$$
we obtain
\bea
\label{a+}
A_+ &=& C\,\frac{i\pi}{4}\,\left \{ (k |\eta_s|)\,\left [
H_\nu^{(1)}(k|\eta_s|)\,H_\mu^{(2)\prime}(k|\eta_s|) -
H_\nu^{(1)\prime}(k|\eta_s|)\,H_\mu^{(2)}(k|\eta_s|)\right ]+ \right .
\nonumber \\
&& \left . (\gamma-\kappa)\,H_\nu^{(1)}(k |\eta_s|)\,H_\mu^{(2)}(k
|\eta_s|) \right \}\,,\\
\label{a-}
A_- &=& C\,\frac{i\pi}{4}\,\left [ (k |\eta_s|)\,\left [
H_\mu^{(1)}(k|\eta_s|)\,H_\nu^{(1)\prime}(k|\eta_s|) -
H_\mu^{(1)\prime}(k|\eta_s|)\,H_\nu^{(1)}(k|\eta_s|)\right ] +
\right. \nonumber \\
&& \left. +(\kappa -\gamma)\,H_\nu^{(1)}(k |\eta_s|)\,H_\mu^{(1)}(k
|\eta_s|) \right \}\,, \eea
and
\bea
\label{b+}
B_+&=& \frac{i\pi}{4}\,\left \{
\left [ A_+\,H_\mu^{(1)}(k \,|\eta_1|)+A_-\,H_\mu^{(2)}(k \,|\eta_1|)
\right ]
\left [ (k \,\eta_1)\,H_\rho^{(2)\prime}(k \,|\eta_1|) +  (\kappa -
\alpha)\right. \right . \nonumber \\
&& \left . \left . H_\rho^{(2)}(k \,|\eta_1|) \right ]-
(k \,\eta_1)\,\left [ A_+\,H_\mu^{(1)\prime}(k
\,|\eta_1|)+A_-\,H_\mu^{(2)\prime}(k \,|\eta_1|) \right ]\,
H_\rho^{(2)}(k \,|\eta_1|) \right \}\,, \\
\label{b-}
B_-&=& -\frac{i\pi}{4}\,\left \{
\left [ A_+\,H_\mu^{(1)}(k \,|\eta_1|)+A_-\,H_\mu^{(2)}(k \,|\eta_1|)
\right ]
\left [ (k \,\eta_1)\,H_\rho^{(1)\prime}(k \,|\eta_1|) +  (\kappa -
\alpha)\right . \right. \nonumber \\
&& \left . \left . \,H_\rho^{(1)}(k \,|\eta_1|) \right ]- (k
\,\eta_1)\,\left [ A_+\,H_\mu^{(1)\prime}(k \,|\eta_1|)+
A_-\,H_\mu^{(2)\prime}(k \,|\eta_1|) \right ]\,
H_\rho^{(1)}(k \,|\eta_1|) \right \}\,,
\eea
where the prime stands for derivative with respect to the argument of
the Hankel function. From the condition $|C|=1$ we get $|A_+|^2 -
|A_-|^2=1$ and  $|B_+|^2 - |B_-|^2=1$, as needed for generic
Bogoliubov coefficients.

\subsection{``Duality'' of the Bogoliubov coefficients}
\label{sec4.2}
In this section we analyse the behaviour of the Bogoliubov coefficients
$B_-$
and $A_-$ under the ``duality'' transformation $\gamma \rightarrow -\gamma$,
$\kappa \rightarrow -\kappa$ and $\alpha \rightarrow -\alpha$, under which
the pump fields are reversed. We will thus check that, with a careful
choice of the matching conditions, the symmetry that can be shown to be exact in the
Hamiltonian approach \cite{dualitypert} is preserved.

In our context we need the following relations among
Hankel functions (see e.g.~\cite{abra})
\bea
&& H^{(1,2)}_{-\xi}(z) = e^{\pm i \pi \xi}\, H^{(1,2)}_{\xi}(z)\,,\\
&& z\,H^{(1,2)}_{\xi-1}(z) + z\,H^{(1,2)}_{\xi+1}(z) = 2\xi\,
H^{(1,2)}_{\xi}(z) \,,\\
&& z\,\frac{d }{d z}\,H^{(1,2)}_{\xi}(z) + z\,H^{(1,2)}_{\xi}(z) = \xi\,
H^{(1,2)}_{\xi-1}(z) \,. \eea

Independently of the range of frequencies we get
\bea
&& \frac{A_-(\gamma, \kappa)}{A_-(-\gamma,-\kappa)} = \exp \left [
i \pi + \frac{i \pi}{2}\left ( \left |\gamma + \frac{1}{2} \right | +
\left |\kappa + \frac{1}{2} \right |
- \left |\gamma - \frac{1}{2} \right | - \left |\kappa - \frac{1}{2}
\right |\right )  \right ] \,,\\ \label{p2}
&& \frac{B_-(\gamma, \kappa, \alpha)}{B_-(-\gamma,-\kappa, -\alpha)} =
\exp \left [i \pi + \frac{i \pi}{2}\left ( \left |\gamma + \frac{1}{2}
\right | +\left |\alpha + \frac{1}{2} \right |
- \left |\gamma - \frac{1}{2} \right | - \left |\alpha - \frac{1}{2}
\right | \right ) \right] \,.
\eea

There are two important comments to be made on the above formulae.
The first is
that ${B_-(-\gamma, -\kappa, -\alpha)}$ differs from ${B_-(\gamma,\kappa,
\alpha)}$  by just a phase. Hence the spectrum (being proportional to
$|B_-|^2$) is identical for a given pump
field or for its inverse. We stress that this duality property holds
independently of the number and characteristics of the intermediate
phases and thus, as argued in \cite{dualitypert}, is  generally valid.
The second observation is that duality
 depends crucially on having imposed
the continuity of the field and of its derivative on  $\Psi_k$ and
{\it not} on the canonical
field $v_k$. The difference
in imposing continuity of
$\Psi_k$ or of $v_k$
arises from the discontinuous nature of the background itself
(actually of $\dot\varphi$) and from the fact that $\Psi_k$ and $v_k$
 obey  equations containing first and second time-derivatives
 of the pump field, respectively. This gives rise to $\delta$-function
contributions in the  case of $v_k$,  making the requirement of
continuity suspicious for that variable.

One welcome consequence of  ``duality'' is the fact that the
antisymmetric tensor field and the axion
have identical spectra since their pump fields are the inverse of
each other (see below). This must be so since
they are just different descriptions of the same physical degree of freedom.

\subsection{General form of the spectral slopes}
\label{sec4.3}
The parameters $\eta_s$ and $\eta_1$ in the formulae for $B_-$ define
two characteristic
comoving frequencies, $k_s = |\eta_s|^{-1}$, $k_1 = |\eta_1|^{-1}$,
which can be
traded for two proper frequencies $f_s$ and $f_1$
 by the standard relations
\beq
 2 \pi f_s  = \left (\frac{k_s}{a} \right )\,, \quad \quad
2 \pi f_1  = \left (\frac{k_1}{a} \right )\,.
\eeq
\begin{table}
\begin{center}
\begin{tabular}{llcl}
{\sl Exponents} & {\sl Bogoliubov coefficient} & {\sl Leading
contribution}& {\sl  Power of $f$ ($\epsilon_{\Upsilon}$)} \\
\tablerule
$\gamma >1/2, \alpha >1/2, \forall \kappa$ &
${\cal C}_1,{\cal C}_2, {\cal C}_3, {\cal C}_4 \neq 0$ & ${\cal
C}_1$& $1-|\gamma|-|\alpha|$ \\
$\gamma >1/2$, $\alpha <1/2$, $\forall \kappa$ &
${\cal C}_1,{\cal C}_2, {\cal C}_4  \neq 0$,
${\cal C}_3 = 0$ &${\cal C}_1$& $-|\gamma-\alpha|$ \\
$\gamma <1/2$, $\alpha >1/2$, $\forall \kappa$ &
${\cal C}_1, {\cal C}_3, {\cal C}_4  \neq 0$, ${\cal C}_2 = 0$&
${\cal C}_1$&
$-|\gamma-\alpha|$ \\
$\gamma <\alpha$, $\gamma<1/2$, $|\alpha|<1/2$, $\forall \kappa$ &
${\cal C}_1 = 0$, ${\cal C}_2, {\cal C}_3, {\cal C}_4 \neq 0$& ${\cal C}_2$&
$-|\gamma-\alpha|$ \\
$\gamma >\alpha$, $\alpha<1/2$, $|\gamma|< 1/2$, $\forall \kappa$ &
${\cal C}_1 = 0$, ${\cal C}_2, {\cal C}_3, {\cal C}_4 \neq 0$&$ {\cal C}_3$&
$-|\gamma-\alpha|$ \\
$\gamma <-1/2$, $\alpha<-1/2$, $\forall \kappa $ &
${\cal C}_1 = 0$, ${\cal C}_2, {\cal C}_3, {\cal C}_4 \neq 0$& ${\cal
C}_4$& $1
-|\gamma|-|\alpha|$ \\
\end{tabular}
\vskip 0.2 truecm
\caption{\small\sl Power of $f$ in the Bogoliubov coefficient $B_-$
 for $f \ll f_1, f_s$}
\label{tab1}
\end{center}
\end{table}

It is easy to see that the two scenarios for the background,
intermediate dual dilaton
phase and intermediate string phase, lead to $f_s > f_1$ and $f_s <
f_1$, respectively.
In the case  $f \ll f_s , f_1$ (fluctuations that exit
in the dilaton phase and re-enter in the radiation phase),
 which is common to both scenarios,
we can approximate the exact result (\ref{b-}) for the Bogoliubov
coefficient $B_-$ in the following way\footnote{We have used the
following relations for $\nu$ not integer:
$H_\nu^{(1,2)} = (1 \pm i\,\cot \nu \pi)\,J_\nu \mp
\frac{i}{\sin \nu \pi}\,J_{-\nu}$, with
$J_\nu = \left (\frac{z}{2}\right )^\nu\,\sum_{k=0}^{\infty}\frac{(-1)^k}
{k\Gamma(\nu+k+1)}\,\left (\frac{z}{2}\right )^{2k}$,
assuming $\nu \neq 0$.}
$$ B_- = {\cal C}_1 + {\cal C}_2 + {\cal C}_3 + {\cal C}_4 + \cdots$$
where
\bea
{\cal C}_1 &=& {\cal N}_1\,\left (\frac{f}{2f_s} \right )^{-\nu}\,
\left (\frac{f}{2f_1} \right)^{- \rho }\,
\left [ {\cal C}_1^1\,\left ( \frac{f_1}{f_s}\right )^{\mu } +
{\cal C}_1^2\,\left ( \frac{f_s}{f_1}\right )^\mu\, \right ]\,, \\
{\cal C}_2 &=& {\cal N}_2\,\left (\frac{f}{2f_s}\right )^{-\nu}\,\left
(\frac{f}{2f_1}\right)^{\rho }\,
\left [{\cal C}_2^1\,\left (\frac{f_1}{f_s}\right )^{\mu } +
{\cal C}_2^2\,\left (\frac{f_s}{f_1} \right )^{\mu }\, \right ]\,, \\
{\cal C}_3 &=& {\cal N}_2\,\left (\frac{f}{2f_s} \right )^{\nu}\,\left
(\frac{f}{2f_1} \right )^{ - \rho }\,
\left[ {\cal C}_3^1\,\left (\frac{f_1}{f_s}\right)^{\mu } +
{\cal C}_3^2\,\left (\frac{f_s}{f_1}\right )^{\mu }\, \right ]\,,
\eea
\beq
{\cal C}_4 = {\cal N}_4\,\left (\frac{f}{2f_s}\right)^{1 - \nu}\,
\left (\frac{f}{2f_1}\right)^{1 - \rho}\, \left [ {\cal C}_4^1\,
\left (\frac{f_s}{f_1}\right )^{\mu-1 }\,
 + {\cal C}_4^2\,\left (\frac{f_1}{f_s}\right )^{\mu+1} +
{\cal C}_4^3\,\left (\frac{f_1}{f_s}\right )^{\mu-1} +
{\cal C}_4^4\,\left (\frac{f_s}{f_1}\right )^{\mu+1}\, \right ]\,.
\eeq
Since, by their definition (\ref{defmu}), $\nu,\rho > 0$,   ${\cal
C}_1$ gives the leading contribution unless the coefficients appearing
in front of it vanish. Table~\ref{tab1} shows which one of the ${\cal
C}_i$ is
dominant for different choices of the background parameters.

In the case $f_1 \ll f \ll f_s$  (fluctuations that exit in the
dilaton phase
and re-enter in the dual-dilaton phase) we get instead:
\beq
B_- = {\cal D}_1\,\left(\frac{f}{2f_s}\right )^{-\mu  - \nu } +
{\cal D}_2\,\left (\frac{f}{2f_s}\right )^{\mu  - \nu } +
{\cal D}_3\,\left (\frac{f}{2f_s}\right )^{-\mu  + \nu } +
{\cal D}_4\,\left (\frac{f}{2f_s}\right )^{2 - \mu  - \nu }\,.
\label{highf}
\eeq

Table~\ref{tab2} shows the leading contribution to $B_-$ in this case.
The explicit form of the coefficients ${\cal N}_i$, ${\cal C}_i^j$ and
${\cal D}_i$ for both cases is given in the appendix.
\begin{table}
\begin{center}
\begin{tabular}{llcl}
{\sl Exponents} & {\sl Bogoliubov coefficient} & {\sl Leading
contribution}&  {\sl  Power of $f$ ($\epsilon_{\Upsilon}$)}\\
\tablerule
$\gamma >1/2$, $\kappa >1/2$ & ${\cal D}_1, {\cal D}_2, {\cal D}_3,
{\cal D}_4 \neq 0$ & ${\cal D}_1$ & $1 -|\gamma|-|\kappa|$ \\
$\gamma >1/2$, $\kappa <1/2$ & ${\cal D}_1,{\cal D}_2, {\cal D}_4 \neq
0$, ${\cal D}_3 = 0$
& ${\cal D}_1$ & $-|\gamma-\kappa|$ \\
$\gamma <1/2$, $\kappa >1/2$ & ${\cal D}_1, {\cal D}_3, {\cal D}_4
\neq 0$,${\cal D}_2 = 0$ & ${\cal D}_1$ & $-|\gamma-\kappa|$ \\
$\gamma <\kappa$, $\gamma<1/2$, $|\kappa|< 1/2$ &
${\cal D}_1 = 0$, ${\cal D}_2, {\cal D}_3,{\cal D}_4 \neq 0$
& ${\cal D}_2$ &  $-|\gamma-\kappa|$ \\
$\gamma >\kappa$, $\kappa<1/2$, $|\gamma|< 1/2$ &
${\cal D}_1 = 0$, ${\cal D}_2,{\cal D}_3, {\cal D}_4 \neq 0$
& ${\cal D}_3$&  $-|\gamma-\kappa|$ \\
$\gamma <-1/2$, $\kappa<-1/2$ &${\cal D}_1 = 0$,
${\cal D}_2,{\cal D}_3, {\cal D}_4  \neq 0$ &${\cal D}_4$ &
$1 -|\gamma|-|\kappa|$ \\
\end{tabular}
\vskip 0.2 truecm
\caption{\small\sl Power of $f$ in the Bogoliubov coefficient $B_-$
for an intermediate dual-dilaton phase and  $f_1 \ll f \ll f_s$ }
\label{tab2}
\end{center}
\end{table}

Tables~(\ref{tab1} and \ref{tab2}) also show the leading
power of $f$ appearing in
the Bogoliubov coefficient $B_-$, in the two above-mentioned cases,
i.e. re-entry in the dual-dilaton or in
the radiation phase.
We can summarize this behaviour as follows
\bea
\label{summaryBog}
&& |B_-|^2 \sim f^{2-2|\gamma|-2|\alpha|}
\quad \quad \gamma > 1/2 \,, \alpha > 1/2 \,\,\,
\mbox{\rm or}\,\,\, \gamma <-1/2\,,\alpha <-1/2 \,,\\
&& |B_-|^2 \sim f^{-2|\gamma -\alpha|} \quad \quad \quad \mbox{\rm all
other cases}\,.
\label{summaryBog1}
\eea

In the case of an intermediate string phase the Bogoliubov
coefficient for
 $f_s \ll f \ll f_1$ (fluctuations that exit in the string phase
and re-enter in the radiation  phase) is given by Eq. (\ref{highf}) after the
substitution $\gamma \rightarrow \kappa , \kappa \rightarrow \alpha$
(hence $\nu \rightarrow \mu , \mu \rightarrow \rho$).
\begin{table}
\begin{center}
\begin{tabular}{clll}
{\sl Particles ($\Upsilon$)} & {\sl Pump field ($\tilde{a}$)} &
 \multicolumn{2}{c}{\sl Spectral slope $n_\Upsilon$} \\
\tablerule
& &{\sl re-entry dual phase}&  {\sl re-entry radiation phase} \\
{\sl  Gravitons} & $a\,e^{-\varphi/2}$ & 4 & 3 \\
{\sl Axions}  & $a\,e^{\varphi/2}$& $4 -4\left | \frac{\delta}{1-\delta}-
\frac{\theta}{1-\theta}\right |$ &
$4-2 \left | \frac{3}{2}-\frac{2\delta}{1-\delta}\right |$ \\
{\sl Heterotic photons} & $e^{-\varphi/2}$
&$4-2\left |\frac{\delta}{1-\delta}-\frac{\theta}{1-\theta}\right |$
&$4-2 \left | \frac{1}{2}-\frac{\delta}{1-\delta}\right |$ \\
$V_{\mu \nu}$ &$e^{-\varphi/2+\sigma}$ & $4-2\left |
\frac{(\delta-\beta_a)}{1-\delta}- \frac{(\theta-\xi_a)}{1-\theta}\right |$&
$4-2\left | \frac{1}{2}-\frac{(\delta-\beta_a)}{1-\delta}\right |$ \\
$W_{\mu \nu}$ &$e^{-\varphi/2-\sigma}$ & $4-2\left |
\frac{(\delta+\beta_a)}{1-\delta}-
\frac{(\theta+\xi_a)}{1-\theta}\right |$ &
$4-2\left | \frac{1}{2}-
\frac{(\delta+\beta_a)}{1-\delta}\right |$ \\
$ B_{a b} $ & $a\,e^{-\varphi/2-2\sigma}$ & $4-4\left |
\frac{\beta_a}{1-\delta}- \frac{\xi_a}{1-\theta}\right |$&
$3 - \frac{4\beta_a}{1-\delta}$ \\
\end{tabular}
\vskip 0.2truecm
\caption{\small\sl Spectral slopes for an intermediate dual-dilaton
phase in the range $f \ll f_1$ (re-entry in radiation phase) and in the
range $f_1 \ll f \ll f_s$ (re-entry in dual-dilaton phase)} \label{tab3}
\end{center}
\end{table}

The spectrum of fluctuations for a generic field $\Upsilon$ is
\beq
\Omega_\Upsilon = \frac{1}{\rho_c}\,\frac{d \rho_\Upsilon}{d \log f} =
{\cal N}_\Upsilon\,\frac{8 \pi^2}{\rho_c}\,f^4\,|B_-^{\Upsilon}|^2\,,
\eeq
where ${\cal N}_\Upsilon$ is the number of polarization states.
We have found it convenient to use a  ``spectral slope'' parameter
 $n_\Upsilon$  defined by the relation
\beq
n_\Upsilon = \frac{d \log \Omega_{\Upsilon}}{d \log f} = 4 + 2
\epsilon_{\Upsilon} \, ,
\eeq
where $\epsilon_{\Upsilon}$ is the exponent appearing
in the $f$-dependence of $|B_-^{\Upsilon}|$ (see Tables \ref{tab1} and 
\ref{tab2}).
The spectral slope, which is simply related to the
usual spectral index by ${\it slope}=({\it index} -1)$,  is more
convenient to describe
the main property of the spectrum, since its sign tells us
 whether the spectrum is increasing  or decreasing with $f$.
We will  now apply the above general results to various possible
backgrounds and
perturbations occurring in string theory.
\begin{table}
\begin{center}
\begin{tabular}{clll}
{\sl Particles ($\Upsilon$)} & {\sl Pump field ($\tilde{a}$)} &
\multicolumn{2}{c}{\sl Spectral slope $n_\Upsilon$} \\
\tablerule
& &{\sl Exit in dilaton phase}&  {\sl Exit in the string phase} \\
{\sl  Gravitons} & $a\,e^{-\varphi/2}$ & 3&
$ \left\{\begin{array}{ll}
6-2\zeta &  \zeta > \frac{3}{2}\\
2\zeta &  \zeta < \frac{3}{2}
\end{array} \right .$\\
{\sl Axions}  & $a\,e^{\varphi/2}$&
$4-2 \left | \frac{3}{2}-\frac{2\delta}{1-\delta}\right |$& $-2\zeta$\\
{\sl Heterotic photons} & $e^{-\varphi/2}$
&$4-2 \left | \frac{1}{2}-\frac{\delta}{1-\delta}\right |$& $4-2\zeta$\\
$V_{\mu \nu}$ &$e^{-\varphi/2+\sigma}$ & $4-2\left | \frac{1}{2}-
\frac{(\delta-\beta_a)}{1-\delta}\right |$ &$4-2\zeta$ \\
$W_{\mu \nu}$ &$e^{-\varphi/2-\sigma}$ & $4-2\left | \frac{1}{2}-
\frac{(\delta+\beta_a)}{1-\delta}\right |$ &
$ 4-2\zeta$  \\
$ B_{a b} $ & $a\,e^{-\varphi/2-2\sigma}$ &
$3 - \frac{4 \beta_a}{1-\delta}$&
$ \left \{\begin{array}{ll}
6-2\zeta &  \zeta > \frac{3}{2}\\
2\zeta &  \zeta < \frac{3}{2}
\end{array} \right .$\\
\end{tabular}
\vskip 0.2 truecm
\caption{\small\sl Spectral slopes for an intermediate string phase
in the frequency ranges $f \ll f_s$ (exit in dilaton phase) and $f_s \ll f \ll
f_1$ (exit during the string phase)}
\label{tab4}
\end{center}
\end{table}

\section{Application to our specific situations}
\label{sec5}
 We now discuss the explicit form of the Schr\"{o}dinger-like equation
(\ref{sch}) for  the fields occurring in the action
(\ref{redaction}). This amounts to finding,
for each perturbation, the relevant pump field and  canonical
variable.
For gravitons and dilatons we refer to~\cite{tensor,scalar}.
For $V_\mu$ and $W_\mu$ the equations of motion are
\bea
\label{a1}
\pa_\mu \left ( \sqrt{-g}\,e^{-\varphi}\,
e^{2\sigma_a}\,V^{\mu \nu a}\right ) &=& 0 \,,\\
\label{a2}
\pa_\mu \left ( \sqrt{-g}\,e^{-\varphi}\,
e^{-2\sigma_a}\,W^{\mu \nu a}\right ) &=& 0\,.
\eea
Using the gauge $W_{0 a}=0=V_{0 a}$
eliminates one unphysical degree
of freedom. However, since the dilaton depends only on time, we can use
the equations of motion to further require
$\nabla \cdot \vec{V_a} = 0, \nabla \cdot \vec{W_a} = 0$. The
equations for the vector fields then take the form (\ref{lag}) and
their  canonical variables are simply:
\bea
&& \psi^j_{V a} = V^j_a\,e^{-\varphi/2+\sigma_a} \,,\\
&& \psi^j_{W a} = W^j_a\,e^{-\varphi/2-\sigma_a} \,.
\eea
The same procedure has been applied for heterotic photons in~\cite{em}.
For the axion field the equation of perturbations around the zero field
solution is~\cite{Copeland}
\beq
A_k^{\prime \prime} + 2 \frac{a^\prime}{a}\,A_k^\prime +
\varphi^\prime\,A_k^\prime + k^2\,A_k = 0\; ,
\eeq
and the canonical variable therefore  is $v_k = e^{\varphi/2}\,a\,A_k$.
It is straightforward to obtain the equation of perturbations for the
 $B_{\mu\nu}$-field
and its canonical variable, i.e.  $v_k = e^{-\varphi/2}\,a^{-1}\,B_k$.

Note that, since the pump fields of the axion and the
antisymmetric tensor
are duality related, the spectrum of their fluctuations will be
the same. For the internal $B$-field we get instead $v_k =
a\,e^{-\varphi/2-2\sigma}\,B_k$.
A list of all relevant ``pump" fields can be found in Tables \ref{tab3} and \ref{tab4}.
In the first  we give the spectral slope for various
fluctuations in the case of an intermediate dual-dilaton phase. The
same is done in Table IV for an intermediate string phase.
We now turn to discussing perturbations in the two scenarios.

\subsection{Intermediate dual-dilaton phase}
\label{sec5.1}
In this scenario, the super-inflationary phase ends at time $\eta=\eta_s$.
Since we assume  such a phase to have washed out any initial spatial
curvature,
the energy density must always be critical. At $\eta=\eta_s$
the dominant source of energy is  the kinetic energy of the
dilaton,  $\rho_\varphi(\eta_s) \sim M_s^4\,e^{-\varphi_s}$.
\begin{figure}
\centerline{\epsfig{file=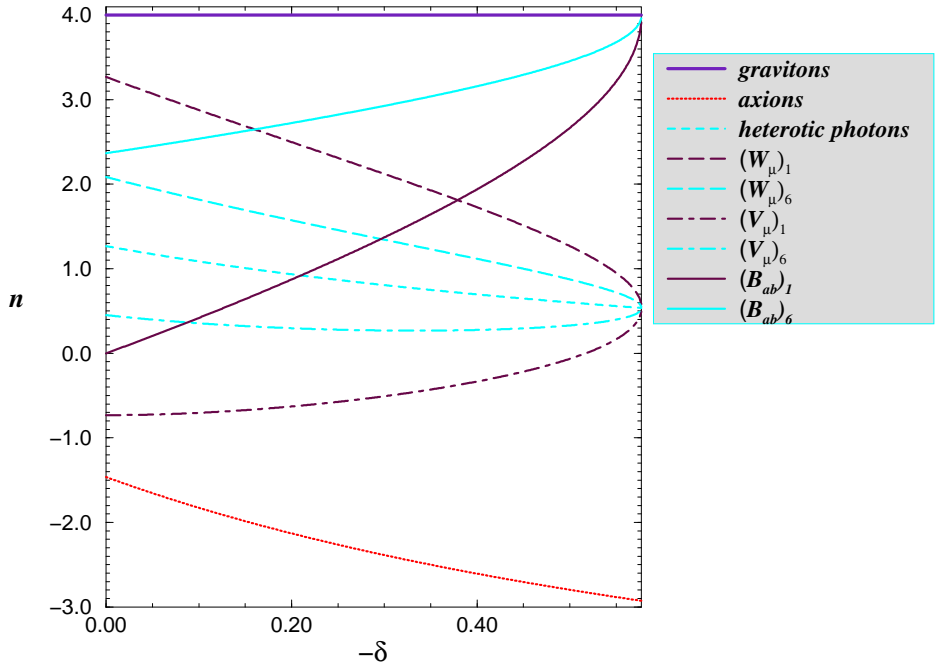,width=0.65\textwidth,angle=-90}}
\caption{\sl Spectral slopes for fluctuations that re-enter in the
dual-dilaton phase when the latter is characterized by $\theta =
1/\sqrt{3}$ (i.e. by constant moduli).}
\label{fig1}
\end{figure}

\begin{figure}
\centerline{\epsfig{file=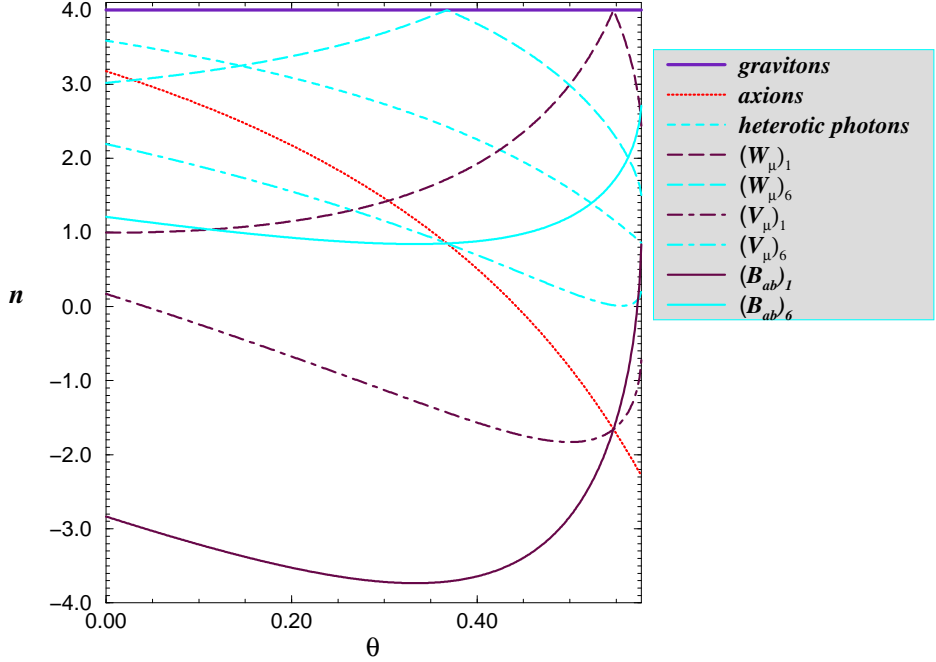,width=0.65\textwidth,angle=-90}}
\caption{\sl Same as Fig.~\ref{fig1} where, instead of fixing $\theta$, we
fix  $\delta =-0.26$ in the intermediate phase.}
\label{fig2}
\end{figure}

\begin{figure}
\centerline{\epsfig{file=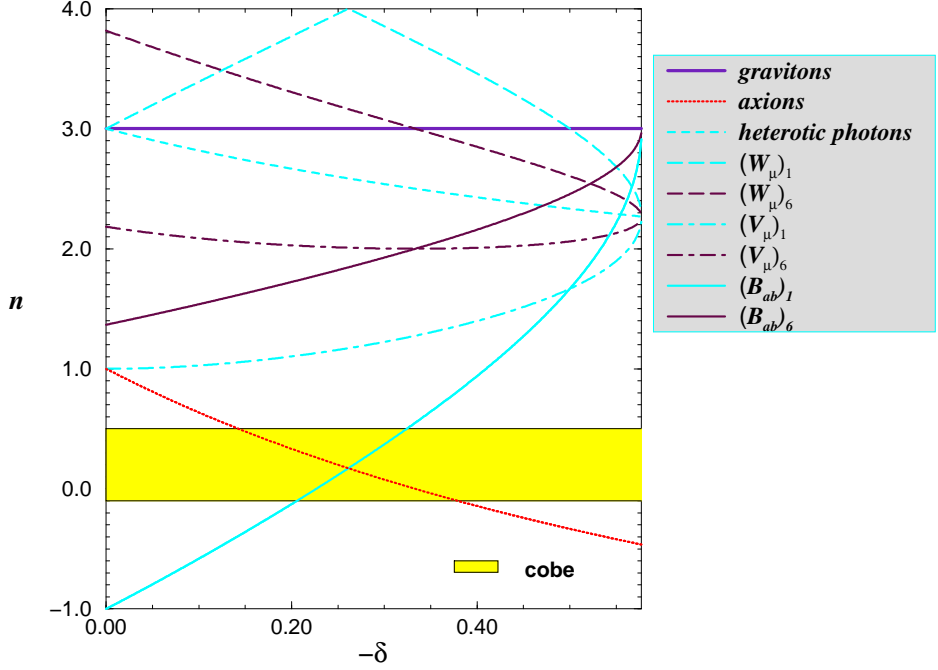,width=0.65\textwidth,angle=-90}}
\caption{\sl Spectral slopes for fluctuations that re-enter in the
radiation phase. For comparison we also show the experimental
constraint from COBE $[20]$: $n + 1 = 1.2 \pm 0.3$.}
\label{fig3}
\end{figure}

\begin{figure}
\centerline{\epsfig{file=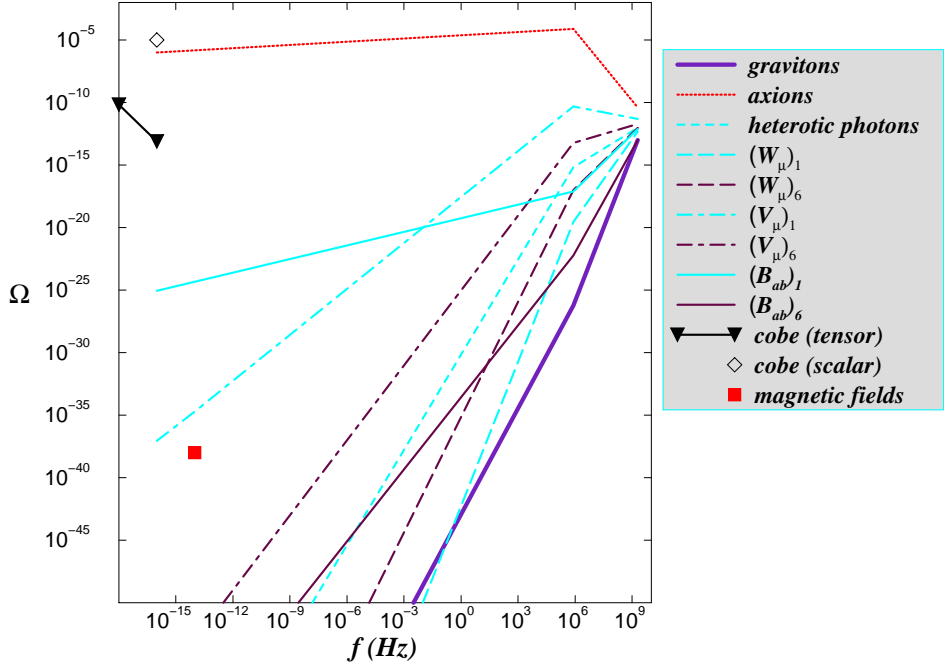,width=0.65\textwidth,angle=-90}}
\caption{\sl Spectrum for all the fluctuations considered in the case
of a dual-dilaton intermediate phase, with the following choice:
$\delta = -0.3$,  \, $\theta = 1/\sqrt{3}$, \,
$f_1 \sim 10^5\,{\rm Hz}, f_s \sim 10^{9}\,{\rm Hz}$,
\,$ \varphi_s \sim -23, \, \varphi_1 \sim -11,$ \,  $T_{\rm
eff}=10^{15}\,{\rm GeV}$.}
\label{fig4}
\end{figure}

Consider now the energy stored in the amplified perturbations during
the dual-dilaton phase. Figs. \ref{fig1}, \ref{fig2} and \ref{fig3} 
give the spectral slopes
in various cases for the two relevant frequency ranges. Fig.~\ref{fig4} gives the
normalization of the spectra in the whole frequency range for
the particular case $\theta = 1/\sqrt{3} , \delta = - 0.3$.
Since all perturbations are of the same order
at the maximal
amplified frequency (here $f_s$), perturbations with (the most)
negative spectral slope
dominate over all others. From the above mentioned figures we see
that the spectral
slope of  axionic fluctuations
re-entering during the dual dilaton phase is generally the most
negative one (at least if we consider isotropic compactifications):
we  thus ignore contributions to the energy density from perturbations
other than the axion's.
The basic idea is to assume that the transition from the dual-dilaton
phase to
the radiation phase occurs precisely when the energy density in
the perturbations becomes critical and starts
to dominate over the kinetic energy of the
coherent dilaton field.

Let us fix for simplicity $\theta = 1/\sqrt{3}$ (i.e. frozen internal
dimensions in
the dual-dilaton phase) and  then impose
criticality  at the end of the dual-dilaton phase in the form:
\beq
M_s^2\,H_1^2 \simeq e^{\varphi_1}\,\rho_A({\eta_1})\,.
\label{crit}
\eeq
Using the equations of motion and assuming $|\eta_1| \gg |\eta_s|$, we have
\beq
H_1 \simeq M_s\,\left |\frac{\eta_s}{\eta_1}\right
|^{(3+\sqrt{3})/2}\,,\quad \quad
e^{\varphi_1} \simeq e^{\varphi_s}\,\left |\frac{\eta_s}{\eta_1}\right
|^{-\sqrt{3}}\,. \label{eq1}
\eeq
Taking into account the results of Sec.~\ref{sec4} we get, apart
from  factors $O(1)$,
\bea
\label{eq2}
\rho_A(\eta_1) && \simeq f_s^4(\eta_s)\,\left (\frac{a_s}{a_1}\right )^4\,
 \left ( \frac{f_1}{f_s} \right )^{n_A}  \simeq
f_s^4(\eta_s)\,\left |\frac{\eta_s}{\eta_1}\right |^{4/(1-\delta)}
\quad  , \,
n_A < 0\,,
\eea
where we have restricted ourselves to the case $n_A < 0$ for the
reasons explained above.

The dependence of the value of the dilaton background at
$\eta=\eta_1$  on the parameter
$\delta$
is completely fixed by the criticality condition Eq.~(\ref{crit}). Indeed,
inserting  Eqs.~(\ref{eq1}) and (\ref{eq2}) in Eq.~(\ref{crit}), we obtain
\beq
e^{\varphi_1} \simeq \left |\frac{\eta_s}{\eta_1}\right
|^{3+\sqrt{3}-4/(1-\delta)}\,. \eeq
If we define an effective temperature $T_{\rm eff}$ at the beginning
of the radiation era by
\beq
\rho_A(\eta_1) \simeq T_{\rm eff}^4 \,,
\eeq
we get
\beq
\label{ratio}
\left |\frac{\eta_s}{\eta_1}\right | \simeq \left (\frac{T_{\rm eff}}{M_s}
\right )^{1-\delta}\,, \eeq
where we have assumed $f_s(\eta_s) \sim M_s$, and thus
\beq
e^{\varphi_1} \simeq  \left (\frac{T_{\rm eff}}{M_s}\right
)^{(1-\delta)(3+\sqrt{3})-4}\,.
\eeq
It is important to stress that this effective temperature may have
nothing to do
with the actual temperature
of a relativistic gas in thermal equilibrium at $\eta = \eta_1$. In
particular, if the coupling
is still very small, axions may not thermalize at all, in spite of
dominating the energy
and of driving a radiation-dominated era. For the same reason,
the fact that $T_{\rm eff}$ can be large in string units should not be
a matter of concern.

Let us now estimate the value of the frequencies $f_s$ and $f_1$ at
present time.  If we assume that the CMB  photons we observe today
carry the (red-shifted) energy of the primordial perturbations
(in particular from axion decay), we have
\bea
&& 2 \pi f_1 =  \frac{k_1}{a_0} \simeq \frac{a_1}{a_0}\,H_1 \,, \\
&& \Omega_\gamma(t_0) = \frac{1}{G(\eta_1) M_{\rm Pl}^2}\,
\frac{H_1^2}{H_0^2}\,\left (\frac{a_1}{a_0} \right )^4\,, \\\vrsmall
&& f_1^4 \simeq G(\eta_1)\,M_{\rm
Pl}^2\,H_0^2\,H_1^2\,\Omega_\gamma(t_0) \simeq
e^{\varphi_1} \,H_0^2\,H_1^2\,\Omega_\gamma(t_0) M_{\rm  Pl}^2/M_{\rm
s}^2\, ,
\label{k2}
\eea
where $H_0$ and $\Omega_\gamma(t_0)$ ($\sim 10^{-4}$) are respectively
the Hubble parameter
and the fraction of the critical energy stored in radiation at the
present time $t_0$.
Using Eq.~(\ref{eq1}) and Eq.~(\ref{ratio}) we finally get
\bea
&& f_1 \simeq \sqrt{H_0 M_{\rm Pl}}
\,e^{\varphi_1/4}\,(\Omega_\gamma(t_0))^{1/4}\,
\left (\frac{T_{\rm eff}}{M_s}\right )^{(1-\delta)(3+\sqrt{3})/4}\,,\\
&& f_s \simeq f_1\,\left (\frac{T_{\rm eff}}{M_s}\right )^{\delta-1}\,.
\eea
Note that, if we choose $T_{\rm eff}=10^{15}\,{\rm GeV}$, corresponding
to a relatively short dual-dilaton phase, and we fix $\delta = -0.3$ in order to
have an almost flat axion spectrum in the low-frequency region
(see Fig.~\ref{fig3}), we get
$f_1 \sim 10^5\,{\rm Hz}, f_s \sim 10^{9}\,{\rm Hz}$ and
$ \varphi_s \simeq -23, \varphi_1 \simeq -11$.
Therefore, the value of the dilaton at the beginning of the radiation
era is still far
from the present value ($\varphi_0 \sim -1$).

Typical spectra for all the fields we have considered are shown in
Fig~\ref{fig4}.
In particular, for  axionic fluctuations
that  re-enter in the dual-dilaton phase ($f_1
\ll f \ll f_s$), we get a decreasing spectrum
\beq
\Omega_A \simeq G(\eta_1)\,H_1^2\,\Omega_\gamma(t_0)\,
\left ( \frac{f}{f_s} \right )^{n_A}\,\left ( \frac{f_s}{f_1} \right
)^{4}\,.
\eeq

On the contrary, for fluctuations that
re-enter in the radiation phase ($f \ll f_1 \ll f_s$), we get
\bea
\Omega_A &\simeq& G(\eta_1)\,H_1^2\,\Omega_\gamma(t_0)\,
\left ( \frac{f}{f_1} \right )^{n_A}\,\left (
\frac{f_s}{f_1} \right )^{2(\nu +\mu)}\,, \\
&\simeq&  \Omega_\gamma(t_0)\,\left (
\frac{f}{f_1} \right )^{n_A}\,,
\eea
which includes the possibility of a scale-invariant flat spectrum.

As can be seen from Fig.~\ref{fig4},  the Kaluza-Klein ``photons"
$V_\mu^{\,\,a}$, can give sufficiently large
seeds for galactic magnetic fields ($\Omega>10^{-38}$ for $f_M\approx
10^{-14}$ Hz~\cite{magn}) in this case, provided of course that the true
electromagnetic field has a non-vanishing component along this direction
in group space. Amusingly enough, this can be achieved in a range of moduli
where axionic perturbations have a nearly flat spectrum.

\subsection{Intermediate string phase}
\label{sec5.2}
In this scenario the Bogoliubov coefficients
are still expressed by Eqs.~(\ref{a+}), (\ref{a-}), (\ref{b+}) and
(\ref{b-}),
$\eta_s$ is the time at which the string phase starts,
and we again assume that the radiation phase, dominated
by the energy stored in the amplified vacuum fluctuations, begins at
$\eta = \eta_1$.
We recall that, in this case, $f_1 > f_s$ and that we expect
$f_1(\eta_1) \sim M_s$.
Since axions  have  the most negative spectral slope,
we impose again that their energy density becomes critical at the
beginning of the
radiation phase:
\beq
H_1^2 = G(\eta_1)\,\rho_A(\eta_1)\,,
\eeq
Using then
\beq
\rho_A(\eta_1) \simeq f_1^4(\eta_1)\,
\left ( \frac{f_s}{f_1} \right )^{n_A}
\eeq
and assuming again that the photons we observe today originate from
the amplified vacuum fluctuations, we can fix the present value of
$f_1$ to be
\beq
f_1(t_0) \simeq \sqrt{H_0 M_{\rm Pl}} \,e^{\varphi_1/4}\,
(\Omega_\gamma(t_0))^{1/4}\,,
\eeq
and relate $\varphi_1$ to the duration of the string phase $z_s=a_1/a_s $
(a free parameter)
\beq
e^{\varphi_1} \simeq z_s^{-2 \zeta}\,.
\eeq
If we define again
\beq
\rho_A(\eta_1) \simeq T_{\rm eff}^4 \,,
\eeq
we obtain
\beq
z_s \simeq \left ( \frac{T_{\rm eff}}{M_s} \right)^{2/\zeta}\,.
\eeq
With the choice $T_{\rm eff}=2\times 10^{18}\,{\rm GeV},\zeta= 0.08 $,
corresponding
to a very long string phase, and fixing $\delta = -1/3$ in order to
have a flat axion spectrum in the low-frequency region
(see Fig.~\ref{fig3}),
we obtain
$$
f_1 \sim 10^{10}\,{\rm Hz}, \,\, f_s \sim 8 \cdot 10^{-1}\,{\rm Hz} \,\, ,
\varphi_s \simeq -8,\; \varphi_1 \simeq -5 \; .
$$
As in the scenario with a dual-dilaton intermediate phase,
we find  the unpleasant result
that the dilaton is still far from its present value
at the beginning of the radiation era.
\begin{figure}
\centerline{\epsfig{file=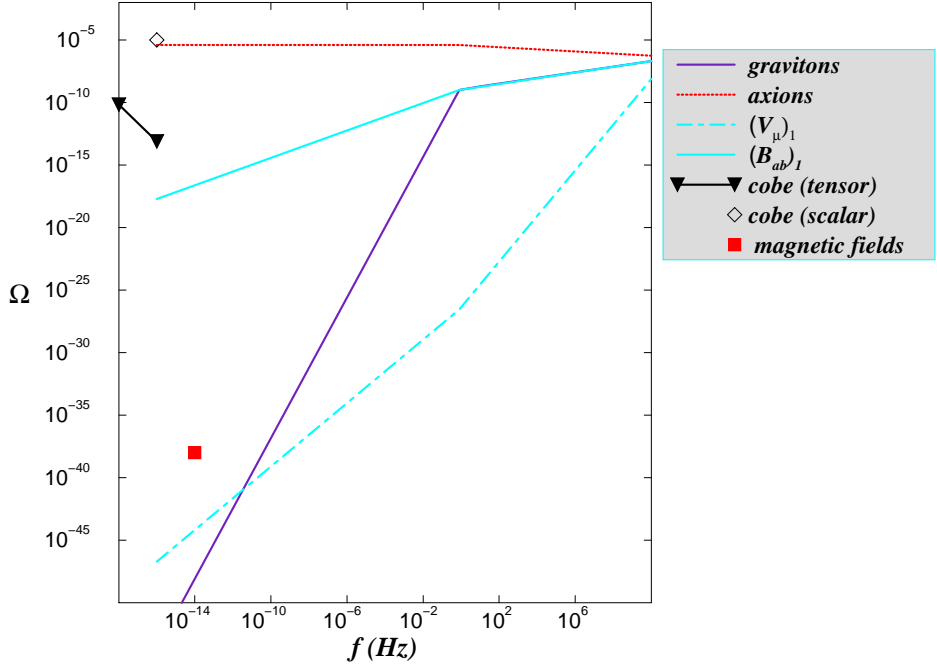,width=0.65\textwidth,angle=-90}}
\caption{\sl Some fluctuation spectra in the case
of a string intermediate phase. The following choice of  parameters
was made: $\zeta= 0.08 $, $\delta = -1/3$,
$f_1 \sim 10^{10}\,{\rm Hz}, f_s \sim 8 \cdot 10^{-1}\,{\rm Hz}$,
$ \varphi_s \sim -8, \varphi_1 \sim -5$, \, $T_{\rm eff}=2 \times
10^{18}\,{\rm GeV}$.}
\label{fig5}
\end{figure}
In Fig.~\ref{fig5} we summarize the results of the spectra for some
perturbations.
The spectrum of the fluctuations that exit in the dilaton phase
 is given in the limit $ f \ll f_s \ll f_1$,
using the coefficients ${\cal C}_i$, shown in
Table~\ref{tab1}.
For  fluctuations that exit in the string phase
 we consider instead the limit $f_s \ll f \ll f_1$,
and the Bogoliubov coefficients are expressed by the quantities ${\cal
D}_i$ (see Table~\ref{tab2}
after substituting $ \gamma \rightarrow \kappa$ and $\kappa
\rightarrow \alpha$).

Note that, for  fluctuations of the axion field
that exit in the string phase, we have
\bea
\Omega_A &\simeq& G(\eta_1)\,H_1^2\,\Omega_\gamma(t_0)\,
\left ( \frac{f}{f_s} \right )^{n_A}\,\left ( \frac{f_1}{f_s}
\right )^{2(\mu +\rho)-4}\,, \nonumber \\
&\simeq & \Omega_\gamma(t_0)\,\left ( \frac{f}{f_s} \right )^{n_A}\,.
\eea
Substituting the parameters of Table~\ref{tab1} we get a decreasing spectrum
\beq
\Omega_A \simeq G(\eta_1)\,H_1^2\,\Omega_\gamma(t_0)\,
\left ( \frac{f}{f_1} \right )^{-2\zeta}\,.
\eeq

In this example, a long string phase produces a
gravitational spectrum of order $\sim 10^{-9}$ in the range of
detection of LIGO/VIRGO,
but a very steep spectrum of Kaluza-Klein photons $V_\mu^{\,\,a}$ at
high frequencies
and consequently a value of perturbations at $f_M \sim 10^{-14}\,{\rm Hz}$
far below the lower limit necessary to seed the dynamo mechanism
for galactic magnetic fields~\cite{magn}.

\section{Discussion}
\label{sec6}
Our main results can be summarised as follows:
Eqs.~(\ref{summaryBog}), (\ref{summaryBog1}),
Tables~\ref{tab3} and ~\ref{tab4},  Figs.~\ref{fig1}, \ref{fig2}
and \ref{fig3} give our main conclusions concerning the spectral slopes
of the various spectra in the two scenarios, while
Figs.~\ref{fig4} and \ref{fig5} illustrate  the
spectra of all perturbations
for certain typical choices of the background's moduli.
Rather than discussing the fine details, we would like to
draw some conclusions, which appear to be relatively robust
with respect to (slight?) variations of the moduli.

\begin{list}{$\diamondsuit$}{\setlength{\leftmargin}{0in} 
\setlength{\rightmargin}{0in}} 
\item Our calculations are based on the use of the low-energy effective action
both for the backgrounds and for the perturbations. Since
in the pre-big bang scenario a high-curvature phase
is  necessary before any exit to standard cosmology can be achieved,
such a procedure is often criticized (see e.g. \cite{Hawking}) and
requires some justification. We have seen in our explicit computations
that the spectrum of long-wavelength perturbations, which exit and re-enter
at small curvatures (in string units), does not depend on the details of the
high curvature phase. Also, the use of higher-derivative-corrected
perturbation equations has recently  been shown \cite{Maurizio} not
to change the
low-frequency spectra
by more than a number $O(1)$.
Thus predictions for the low-energy end of the spectra
appear to be robust. Why? The physical explanation
almost certainly lies in the freezing-out of super-horizon-scale
perturbations. The occurrence of a constant mode at
sufficiently large wavelengths
can be shown without  reference to the low-energy approximation
\cite{dualitypert}
and, by a canonical transformation argument, should also apply to
the constant-momentum mode.
By contrast, the high frequency spectra are expected to depend
quite crucially on the details of the strong curvature transition.
We expect our naive formulae to give  ``lower bounds" for those
parts of the spectra.
\item The main result of our investigation is  the confirmation of
the suggestion found in ref.\cite{Copeland} that positive spectral slopes
are by no means a must in pre-big bang cosmology. By computing the spectra
 after re-entry, we have confirmed that axions do have, more often than not,
negative slopes (decreasing spectra). However, other fields, such as
KK gauge fields and scalars, can also exhibit negative slopes. A
particularly promising
case is the one shown in Fig.~\ref{fig3}, since, in a region around the one with
nine-dimensional symmetry ($\delta = \pm \beta_a = - 1/3$),
the axion spectrum
and that of a KK scalar field are nearly flat, while the slope of the
spectrum of some KK gauge field
is positive but sufficiently small to produce large enough seeds
for the galactic   magnetic fields.
\item
Unfortunately, the promising results of Fig.~\ref{fig3} are somewhat spoiled
when a long
dual-dilaton or string phase is inserted in the background between
the dilaton
and radiation phases. In this case, the spectral slopes grow somewhat wild
in the negative direction (see e.g. Fig.~\ref{fig2} for the dual-dilaton
case), making some of the spectra peak at  very low frequency.
The generic consequence of this phenomenon is a huge increase in the
total, integrated energy density in the perturbations. If the string
coupling
(the dilaton) is not very small throughout the intermediate phase,
the energy
in these perturbations soon becomes  critical and the intermediate
phase stops.
The only way to have a long intermediate phase is therefore to force
the dilaton to be very
perturbative until the end of the intermediate phase, be it the
string or the dual-dilaton
phase. In this case, however, at the beginning of the radiation phase
 the dilaton
is still very much displaced from its present value
(where supposedly the minimum of its non-perturbative potential is)
and may have a hard time reaching it later.
In other words, the most appealing possible scenarios appear
 to be those with a sudden transition
between the dilaton and radiation phases occurring at ``realistic" values
 for the string coupling
(roughly $1/N$, if $N$ is the number of  effectively amplified
distinct species).
Although this appears at present as some kind of fine tuning
of the ratio of two moduli,
it is not excluded that a better understanding of the initial conditions
leading to PBB behaviour along the lines of Ref. \cite{BMUV} may tie
together the initial values of the coupling and the curvature so that
such conditions are naturally realised.

\item If  the latter picture is adopted, it is possible
to have a nearly scale-invariant dilaton/moduli spectrum.
This could lead to an interesting mechanism to generate large-scale
anisotropy
along the lines given in Ref.\cite{em1}. In the same region of moduli space
one obtains reasonably
large  fluctuations of the KK gauge fields to provide sizeable  seeds
 for the galactic magnetic fields. On the negative side,
in this
region of parameter space, the situation would
be quite discouraging for
generating a large enough gravitational-wave signal in the
interesting frequency range.
\end{list}

\section*{Note added}
While completing this work we  became aware of a paper by
Brustein and Hadad
\cite{BH} which is also dealing with generic perturbations in string
cosmology.
Their method is somewhat different from ours: instead of working
within a specific parametrization of the high-curvature phase, they
have assumed the freezing
of the fluctuation and of its conjugate
momentum for super-horizon scales. Also, they have
not imposed our
criticality condition and thus have not obtained predictions on the
value of the dilaton
at the beginning of the radiation phase. We have checked
that our results agree with theirs wherever a comparison is possible.

\section*{Acknowledgements}
This work was supported in part by the EC contract No. ERBCHRX-CT94-0488.
A. B. and C. U. are partially supported by the University of Pisa.
We acknowledge useful discussions with B. Allen, J.D. Barrow, M. Bruni,
T. Damour, G. Dvali, F. Finelli, M. Gasperini, M. Maggiore and
V.F. Mukhanov. We are particularly grateful  to  R. Brustein
for interesting exchanges of information
on his related work and for interesting discussions.

\appendix
\section{}
\label{app}
Here we give the explicit form of the coefficients
entering the  Bogoliubov formulae of Sec.~\ref{sec4.3}.
In the case of the limit $f \ll f_1 \ll f_s$ we get:
\bea
{\cal N}_1 &=& \frac{i}{8\pi}\,\frac{\Gamma(\mu)\,\Gamma
(\nu)\,\Gamma(\rho )}
{\Gamma(1 + \mu )}\,,\nonumber \\
{\cal C}_1^1 &=& \left( \gamma  - \kappa  + \mu  + \nu  \right ) \,
\left( -\alpha  + \kappa  + \mu  - \rho  \right )\,,\nonumber \\
{\cal C}_1^2 &=& \left( \gamma  - \kappa
- \mu  + \nu  \right )\,\left( \alpha  - \kappa  + \mu  + \rho
\right)\,; \nonumber
\eea
\bea
{\cal N}_2 &=& -\frac{i}{8\pi}\,\frac{\Gamma(\mu)\,\Gamma
(\nu)\,\Gamma(-\rho )}
{\Gamma(1 + \mu )}\,\left [\cos (\pi \,\rho ) - i\,\sin (\pi \,\rho )
\right ] \,,\nonumber \\
{\cal C}_2^1 &=& \left( \gamma  - \kappa  + \mu  + \nu  \right ) \,
\left( \alpha  - \kappa  - \mu  - \rho  \right )\,,\nonumber \\
{\cal C}_2^2 &=& \left( \gamma  - \kappa
- \mu  + \nu  \right )\,\left( -\alpha  + \kappa  - \mu  + \rho
\right)\,; \nonumber
\eea
\bea
{\cal N}_3 &=& -\frac{i}{8\pi}\,\frac{\Gamma(\mu)\,\Gamma
(-\nu)\,\Gamma(\rho )}
{\Gamma(1 + \mu )}\,\left [\cos (\pi \,\nu ) - i\,\sin (\pi \,\nu )
\right ] \,,\nonumber \\
{\cal C}_3^1 &=& \left( \gamma  - \kappa  + \mu  - \nu  \right ) \,
\left( \alpha  - \kappa  - \mu  + \rho  \right )\,,\nonumber \\
{\cal C}_3^2 &=& \left( -\gamma  + \kappa
+ \mu  + \nu  \right )\,\left( \alpha  - \kappa  + \mu  + \rho
\right)\,;  \nonumber
\eea
\bea
{\cal N}_4 &=&
-\frac{i}{8\pi}\,\frac{\Gamma(\mu)\,\Gamma(\nu)\,\Gamma(\rho)}
{\Gamma(2 - \mu )\,\Gamma(1 + \mu )\,\Gamma(2 + \mu )\,
\Gamma(2 - \nu )\,\Gamma(2 - \rho )}\,,\nonumber \\
{\cal C}_4^1 &=& \left( \alpha  - \kappa  + \mu  + \rho  \right) \,
\Gamma(2 + \mu )\,\Gamma(2 - \rho )\,
\left [ \left( -2 + \gamma  - \kappa  - \mu  + \nu  \right)  \right
.\nonumber \\
&& \left .  \,\Gamma(2 - \mu )\,\Gamma(1 - \nu ) +
\left( 2 + \gamma  - \kappa  - \mu  + \nu  \right) \,\Gamma(1 - \mu
)\,\Gamma(2 - \nu ) \right]\,,\nonumber \\
{\cal C}_4^2 &=& \left( -\alpha  + \kappa  + \mu  - \rho  \right) \,
\Gamma(2 - \mu )\,\Gamma(2 - \rho )\,\left[ \left( -2 + \gamma  -
\kappa  + \mu  + \nu  \right)\right . \nonumber \nonumber \\
&& \left. \,\Gamma(2 + \mu )\,\Gamma(1 - \nu ) +
\left( 2 + \gamma  - \kappa  + \mu  + \nu  \right) \,
\Gamma(1 + \mu )\,\Gamma(2 - \nu ) \right]\,,\nonumber \\
{\cal C}_4^3 &=& \left( \gamma  - \kappa  + \mu  + \nu  \right) \,
\Gamma(2 + \mu )\,\Gamma(2 - \nu )\,
\left[ \left( 2 - \alpha  + \kappa  + \mu  - \rho  \right) \,\right .
\nonumber \\
&& \left.  \Gamma(2 - \mu )\,\Gamma(1 - \rho ) +
\left( -2 - \alpha  + \kappa  + \mu  - \rho  \right) \,
\Gamma(1 - \mu )\,\Gamma(2 - \rho ) \right]  \,,\nonumber \\
{\cal C}_4^4 &=& \left( \gamma  - \kappa  - \mu  + \nu  \right) \,
\Gamma(2 - \mu )\,\Gamma(2 - \nu )\,
\left[ \left( -2 + \alpha  - \kappa  + \mu  + \rho  \right) \right.
\nonumber \\
&& \left . \Gamma(2 + \mu )\,\Gamma(1 - \rho ) +
\left( 2 + \alpha  - \kappa  + \mu  + \rho  \right) \,
\Gamma(1 + \mu )\,\Gamma(2 - \rho ) \right ]\,; \nonumber
\eea
while for fluctuations in the frequency region $f_1 \ll f \ll f_s$ we
obtain:
\bea
{\cal D}_1 &=& -\frac{i}{8}\,e^{-i\pi/2 \,\left( -\mu  + \rho  \right) }\,
\left( -\gamma  + \kappa  + \mu  - \nu  \right) \,\pi^{-1} \,
2{\Gamma}(\mu)\, {\Gamma}(\nu)\,,\nonumber \\
{\cal D}_2 &=& -\frac{i}{8}\,e^{-i\pi/2 \,\left( \mu  + \rho  \right) }
\left( -\gamma  + \kappa  - \mu  - \nu  \right)\,\pi^{-1}\,
2{\Gamma}(-\mu )\,{\Gamma}(\nu ) \,,\nonumber \\
{\cal D}_3 &=& -\frac{i}{8}\,e^{-i\pi/2 \,\left( 2\nu-\mu  +\rho \right) }
\left( -\gamma  + \kappa  + \mu  + \nu  \right)\,\pi^{-1}\,
2\Gamma(\mu)\,{\Gamma}(-\nu ) \,,\nonumber \\
{\cal D}_4 &=& \frac{i}{8}\,e^{-i\pi/2 \,\left( -\mu  + \rho  \right)
}\,\pi \,
\left [ \left( 2 - \gamma  + \kappa  + \mu  - \nu \right) \,
(1 - \mu )\,+ \right .\nonumber \\
&& \left. \left( -2 - \gamma  + \kappa  + \mu  - \nu  \right) \,
(1 - \nu ) \right] \,
\frac{2\Gamma(\mu)\,\Gamma (\nu)} {(1 - \mu )\,(1 - \nu )}\,. \nonumber\eea

\end{document}